\documentstyle[proceedings]{crckapb}
\input{epsf.sty}

\begin{opening}
\title {Analysis of Large Scale Structure using
Percolation, Genus and Shape
Statistics}
\runningtitle {Percolation, Genus and Shape  
Statistics.}
\author {Varun Sahni}
\institute{Inter-University Centre for Astronomy \& Astrophysics,
Post Bag 4, Pune 411007, India}
\end{opening}
\begin{document}

\begin {abstract}

We probe gravitational clustering in N-body simulations using geometrical
descriptors sensitive to `connectedness': 
the genus curve, percolation and shape statistics.
As gravitational clustering advances, the
density field in N-body simulations 
shows an increasingly pronounced departure from Gaussianity
reflected in the changing shape of the percolation curve and the
changing amplitude and shape of the genus curve. We feel that 
both genus and percolation curves provide
complementary probes of large scale structure topology
and could be used to discriminate
between models of structure formation and the analysis of observational
data such as
galaxy catalogs and MBR maps.  The filling factor in
clusters \& superclusters at percolation is small indicating 
that matter 
is more likely to lie in filaments and pancakes. 
An analysis of `shapes' in N-body
simulations has shown that filaments are more pronounced than pancakes. 
To probe shapes of clusters and superclusters more rigorously 
we propose a new shape statistic which does not fit
isodensity surfaces by ellipsoids (as done earlier).
Instead our shape statistic is derived from fundamental 
properties of a compact body such as its
volume $V$, surface area $S$,
integrated mean curvature $C$, and connectivity (characterized by
the Genus). The new shape statistic 
gives sensible results for topologically simple surfaces such as the ellipsoid,
and for more complicated surfaces such as the torus.

[Invited talk, to appear in:
Proceedings of the IAU Symposium No. 183:
``Cosmological parameters and evolution of the Universe'',
Kyoto, Japan Aug. 1997,
ed. K. Sato (Kluwer Academic Publ.)]
\end {abstract}


\section {Introduction}
The Universe as we perceive it seems abundantly rich in visual form.
Its large scale structure consisting of clusters and superclusters of galaxies
has been variously perceived to be 
 `a cosmic web', `network of surfaces', 
`sponge-like', bubble-like' etc. 
Attempts to describe its large scale features quantitatively have been made 
using
a number of statistical indicators sensitive to the `connectedness' of large
scale structure including the genus curve and percolation statistics
\cite{zel82,sh83,gmd86}; minimal spanning trees \cite{bbs85};
Minkowski functionals; 
\cite{mbw94} and
statistics sensitive to shape 
(see Sahni \& Coles (1995) and references therein).

In this talk we assess the relative merits of genus and percolation curves 
by applying them to the same N-body simulations in an $\Omega = 1$ Universe
with scale-invariant initial conditions $P(k) \equiv \langle|\delta_k|^2\rangle
\propto k^n, n = -2, -1, 0, +1$, (for simplicity we show results only
for $n = -2$ which may be considered the lower 
limit of the slope of the initial spectrum on galaxy scales). 
Results are shown at several epochs each
characterized by the scale of nonlinearity, $k_{NL}$, at that epoch 
measured
in units of the fundamental mode $2\pi /L$, 
where $L$ is the length of the simulation box.
N-body simulations were performed on a $128^3$ grid using a particle-mesh algorithm \cite{msh93}. A reduced grid of size $64^3$ was used to construct density
fields from particle positions and the analysis of percolation and genus curves 
was then performed on these fields.

We also introduce a {\it new} statistic sensitive to shape based on
Minkowski functionals. 

\section {Growth of non-Gaussianity during Cosmological gravitational clustering.}
Conventional models of gravitational clustering usually assume
that primordial density perturbations had a scale-invariant Harrison-Zeldovich 
spectrum and were distributed in the manner of a
Gaussian random field.  
Arguments which support this hypothesis stem from the central limit
theorem and the Inflationary paradigm. The non-Gaussianity which we observe 
in the Universe today (clusters, superclusters, voids) 
is attributed to non-linear evolution and the resulting phase correlation between modes.
Two robust and widely used statistical indicators of clustering are the
probability density function (PDF) and the two point correlation function
$\xi(r)$. However neither characterizes the nonlinear distribution of matter
uniquely.
The two point correlation function 
$\xi(x) = \int  d^{3} \vec{k} \exp (i \vec{k} \cdot \vec{x})P(k)$ being
sensitive only to the power spectrum
$P(k) \equiv \langle|\delta_k|^2\rangle$ and not to the phases $\phi_k$ of 
individual modes $[{\delta}(\vec{k})=|\delta_k|\exp(i\phi_k)]$ misses features arising because of phase correlations in the nonlinear regime.
On the other hand,
the PDF does not characterize a distribution uniquely in the nonlinear
regime: distributions with identical PDF's can have very different
topological properties and, conversely, distributions differing in their PDF's 
may have 
identical geometrical properties (this happens for instance in the case
of distributions related by a mapping $\delta_{NL} = F(\delta_{LIN})$, 
such as the log-normal). 

\section {Percolation and Genus Curves}
It is clear that traditional indicators of clustering: the two point correlation
function
and the probability density function, 
must be complemented in the non-linear regime if we are
to get a better understanding of the issue of non-Gaussianity. 
One way of achieving this
is to use geometrical measures which are sensitive to the connectedness
of a distribution. Two such indicators -- percolation and the genus curve,
will be studied in this section,
a third {\it shape statistics}, will be discussed in the next.

One of the aims of percolation theory is to study the connectedness of
structure as a function of the density threshold.
Varying the density threshold from a high to low value, leads to a `percolation
transition' as the volume fraction in the largest cluster changes
rapidly from almost zero to unity when the density
threshold crosses a critical value.
It is convenient to characterize percolation in terms of 
filling factor -- henceforth $FF$ -- 
the total volume in all clusters/voids above/below 
the density contrast threshold divided by the simulation volume
\footnote {$FF$ is the cumulative probability distribution function: 
$FF = P(\delta > \delta_T)$.}.
Gaussian random fields percolate at the critical filling factor
$FF_C \simeq 16\%$ regardless of the spectrum.
Density fields evolving under gravitational
instability typically percolate at lower levels of $FF_C$
depending upon the initial spectrum and the extent of non-linear evolution
\cite {ys96}. Similar conclusions are also reached in the case of point like
distributions although the natural reference in this case
is the Poisson distribution {\cite{ks93}.

Earlier work on gravitational clustering focussed on $FF_C$ as a diagnostic
measure
\cite{sh83,ds92,ks93}.
However, although useful in probing the extent of
nonlinear evolution $FF_C$ does have certain drawbacks, for instance it is
sensitive to resolution, number of particles and sample geometry 
\cite{dw85}.
A powerful new statistic without the above limitations 
is the {\it percolation curve} (PC).
Consider the volume fraction $v_{max}$  defined as the ratio of the volume in
the largest cluster/void to the total volume in all clusters/voids
lying above/below a density contrast threshold. 
The percolation curve 
describes the {\it volume fraction} $v_{max}$ 
as a function of the density contrast
threshold (or filling factor).

\begin {figure}[ht]
\centering
   \begin {minipage}[c]{3in}
\epsfxsize=3in
 \epsfbox{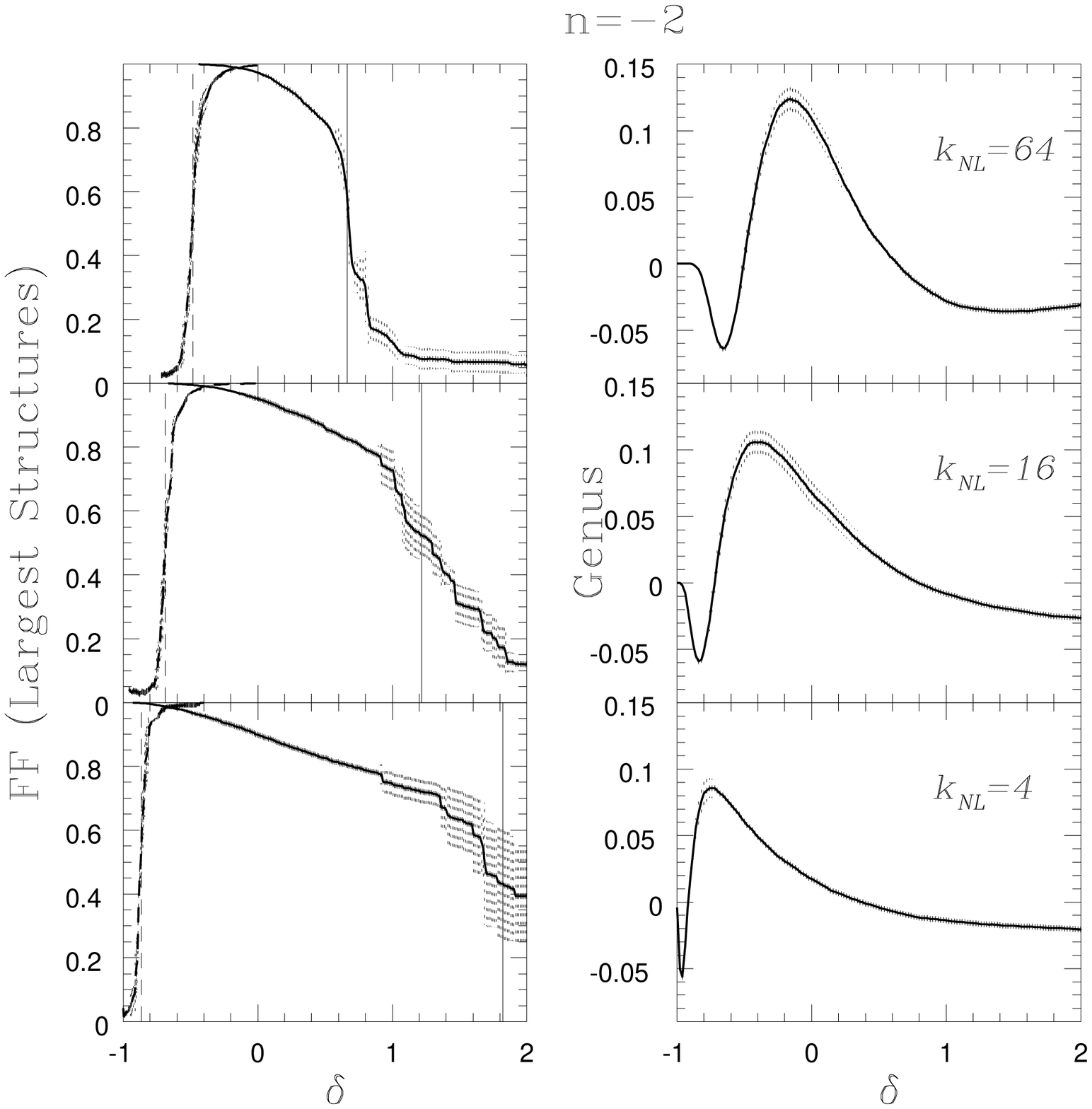}
   \end {minipage}
      \begin {minipage}[c]{3in}
\epsfxsize=3in
     \epsfbox{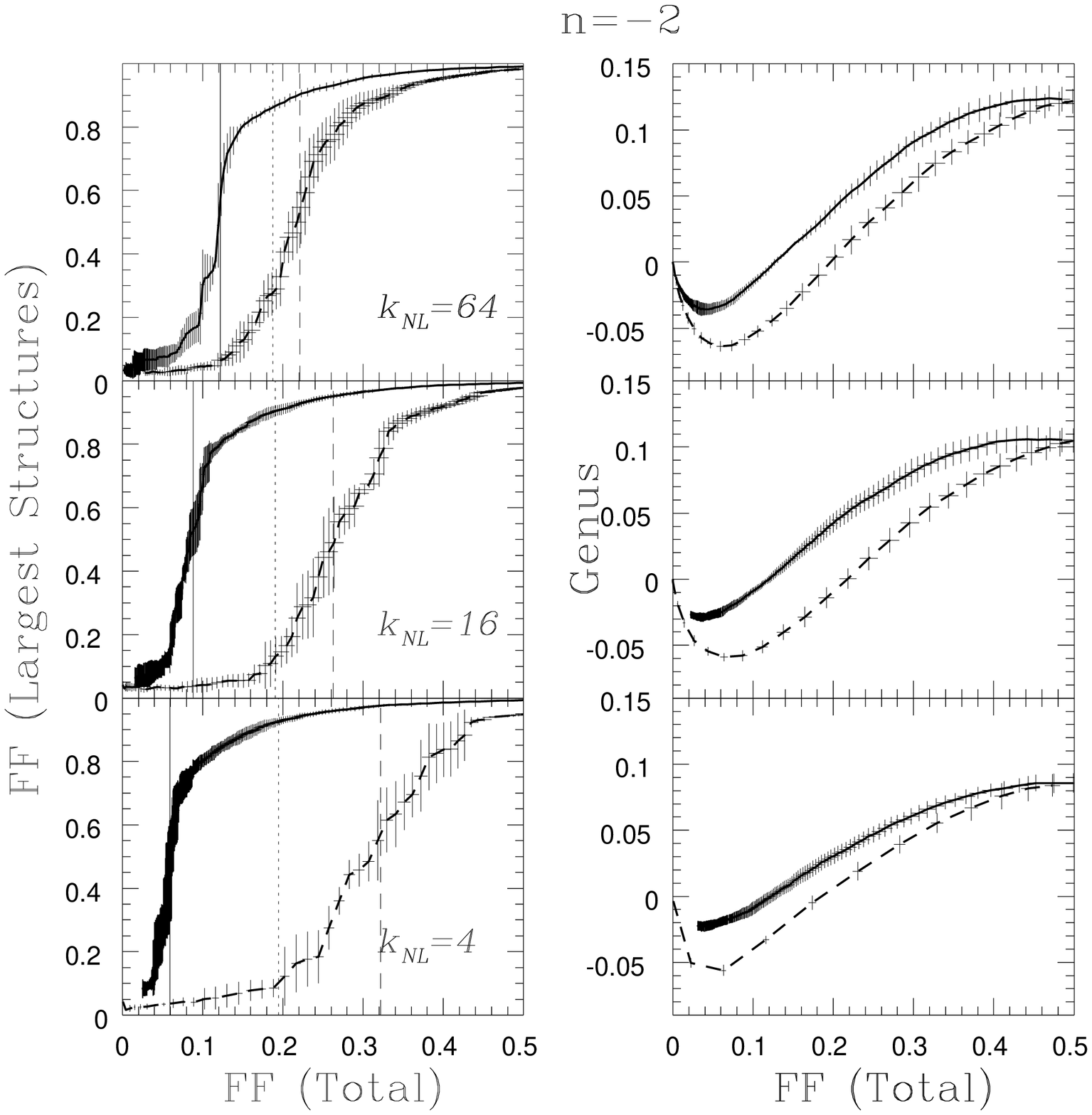}
	\end {minipage}

\caption {\small Percolation (left panels) and genus (right panels) curves
are shown as functions of the density contrast $\delta$ (above) and filling factor (below) for a
scale free initial spectrum $n=-2$. In plots showing PC (left panels),
solid (dashed) curves correspond to 
the volume fraction in the largest cluster (void) -- $v_{max}$. 
Vertical solid (dashed) lines
show the threshold describing percolation between opposite faces of the cube
for clusters (voids). The thin dotted line in the lower left 
hand panel shows the filling factor at percolation for a Gaussian random field 
with the same power spectrum as of evolved density fields
(for details see Sahni, Sathyaprakash \& Shandarin 1997a).}
\end {figure}

The percolation curve is plotted in Fig. 1 for evolved density fields
from N-body simulations with the power-law initial spectrum $n = -2$.
Percolation curves for clusters (thick solid lines) and voids 
(thick dashed lines) are plotted separately. Vertical thin solid and thin dashed
lines show the percolation density threshold below which clusters and 
above which voids percolate. From Fig. 1 we find that at very high (low)
thresholds the number of clusters (voids) is very small and $v_{max} << 1$.
As the threshold is gradually decreased (increased) the volume fraction in 
the largest cluster (void) increases as clusters (voids) begin to merge until 
the percolation transition when the largest `supercluster' (supervoid)
spans the entire simulation box. (Decreasing (increasing) 
the density threshold corresponds
to increasing the filling factor for clusters (voids).)
From Fig. 1  we see that 
as the simulation evolves
$\delta_C$ increases monotonically
as power in longer wavelengths
causes structures to form and align on increasingly larger scales.
For spectra with lesser long range power such as $n=0$, $\delta_C$ initially
increases but later {\it begins to decrease}
signaling the formation of small, isolated clumps (not shown) \cite{sss97a}.

An analysis similar to percolation can also be performed using the 
{\it genus curve} (GC) which can be formally expressed as an integral over the Gaussian curvature $K$
of the iso-density surfaces $S_{\nu}$ lying above/below a density threshold
$\nu=\delta/\sigma_{\delta}$ by the Gauss-Bonnet theorem:
$4\pi G(\nu) = -\int_{S_{\nu}} K d A$.
For Gaussian Random fields the genus curve has a `bell shaped' form:
$G(\nu) = A (1 - \nu^2) \exp(-\nu^2/2)$ \cite{hgw86,gwm87,gott89}.
(An analytical expression for the genus in the weakly non-linear regime has
been obtained in \cite{matsu}.)
Multiply connected surfaces have $G \ge 0$ while simply connected
have $G < 0$. The upper right hand panels of Fig. 1 show the genus curve plotted
as a function of the density contrast. It is interesting to note that
zero-crossings of the genus curve are quite close to 
the percolation threshold for both
clusters and voids. This reflects the fact that the structure transforms 
from simply connected to multiply connected at the zero-crossing of $G$
which allows it to percolate.
One can discern a strong increase in non-Gaussianity 
as the simulation evolves,
reflected by an evolution in shape of both 
percolation and genus curves.

Instead of plotting GC and PC against the density contrast (which is not a 
normalized quantity), it may be more appropriate to plot them against the 
filling factor. This helps to distinguish between distributions related by a 
mapping $\delta \rightarrow f(\delta)$ (such as the log-normal) which have
identical topological properties but can have quite different PDF's.
The lower panels in fig.~ 1 show PC, GC for clusters (solid) and voids (dashed) 
plotted against FF.
The three vertical lines show the filling factor at percolation for
clusters, voids and Gaussian random fields with identical spectra.
Both percolation and genus curves now resemble `hysteresis' curves, the
area between void and cluster curves indicating the degree of non-Gaussianity 
in the distribution. For PC we notice a marked 
increase in non-Gaussianity reflected in the increasing
difference between percolation thresholds for clusters and voids measured by: 
$FF_C(voids) - FF_C(clusters)$. 
The genus curve does not appear to evolve much,
which is surprising. However the amplitude of GC does decrease with epoch, an
effect which is more pronounced for spectra with greater small scale power, 
and which we attribute to the rapid build up of phase correlations due to 
nonlinear mode coupling during
advanced gravitational clustering \cite{sss97a}.


Comparing the geometrical properties of a distribution to a featureless 
Gaussian, one
can make statements regarding its `connectedness or topology'.
In Fig.~1 we have indicated the
percolation threshold of Gaussian random fields by a dotted vertical line.
Comparing these percolation thresholds with those of clusters and
voids we conclude that percolation is
`easier' for clusters and more `difficult' for voids. Clusters percolating
at lower $FF$ than Gaussian are said to possess a
`network-like' topology.  Voids on the other hand percolate at higher
$FF$ than Gaussian and so have a `bubble-like' topology. This appears to
be a generic feature of most models of gravitational clustering with a
reasonable amount of long-wavelength power in the initial spectrum
({\it i.e} $n \le 0$)
\cite{sss97a}. 

\section {Shape-statistics}
As discussed in the last section, gravitating systems clustering
from Gaussian initial conditions percolate at low values
of the filling factor. For CDM \& CHDM models the filling factor can be 
as small as $2\% - 7\%$, much smaller than
the $16\%$ expected for a random Gaussian field \cite{ks93}.
This immediately suggests that the percolating phase is more likely to
be sheet or filament-like since such distributions are likely to occupy a larger linear
dimension (for an equal amount of mass) and will
therefore percolate more easily.
Some indication that this is indeed the case also comes from the Zeldovich 
approximation \cite{sz89}.

A detailed study of `shapes' in scale invariant 
models of gravitational clustering revealed that one dimensional
`filaments' are more abundant
than two-dimensional `sheets'. The filamentarity and pancakeness of 
structures grows with time, leading to the development of a long coherence
length scale in simulations \cite{sss96}. 
Exploring the `connectedness' of large scale structure semi-analytically,
Bond and collaborators recently concluded that clusters and superclusters
appear to be interwoven in a `cosmic web', with superclusters acting 
as cluster-cluster `bridges'.
More pronounced bridges are likely to form between 
clusters of galaxies which are aligned and close together \cite{bkp96}. 

The supercluster-void morphology is 
likely to vary for different scenario's of structure formation,
it is unlikely that structure formation models based on gravitational instability
will have identical topological properties as those
based on string/texture models or explosions.
We therefore feel that a study of shapes of superclusters/voids
could help
distinguish between different alternatives once galaxy catalogues
become fully three dimensional and  a detailed comparison between theory
and observations becomes possible. Forthcoming redshift surveys such as SDSS
and 2dF promise
to shed more light on issues such as whether the `great walls' appearing in
Northern and Southern sky surveys are planar objects or are more like filaments or
`ribbons'.

Most shape statistics proposed so far study the shape of a collection of points
by measuring its moment of inertia tensor, a procedure which is quite similar
to fitting by an ellipsoid. Although this method has yielded some interesting
results it is fair to say that none of the statistics applied to shapes is
entirely satisfactory \cite{sssf97}.
To illustrate this consider two examples: 
(1) the shape of an empty cup as determined from its moment of inertia
tensor 
is approximately ellipsoidal whereas the cup is really a curved 
two-dimensional object.
(The Zeldovich approximation in fact suggests
that the caustic surfaces of the first pancake-like singularities 
are more likely to be curved than flat.)
(2) A torus has both planar and filamentary properties;
fitting with an ellipsoid would suggest an 
oblate shape, whereas a `thin' torus is clearly more like a curved filament.

Results of N-body simulations clearly demonstrate that shapes of
isodensity surfaces vary widely when viewed at different density thresholds.
At high thresholds density peaks are mostly spheroidal, whereas
at closer to percolation thresholds, surfaces get
rather `spongy' with a complicated topology.

To assess the shapes of objects which may be topologically non-trivial,
we have recently proposed a shape statistic based on 
the four Minkowski functionals of a compact surface:
(i) its Volume $V$, (ii) surface area $S$, (iii) integrated mean curvature:
$C = {1\over 2}\int(\kappa_1 + \kappa_2) dS$, and 
(iv) Genus $G$.

From $V$, $S$, $C$ and $G$ we construct three dimensionful and two 
dimensionless shape functions \cite{sss97b}. The dimensionful shape functions 
are: ${\cal H}_1 = V/S$, ${\cal H}_2 = S/C$ and ${\cal H}_3 = C$.
(For multiply connected surfaces $C/G$ may be more appropriate than
$C$.) The pair of dimensionless shape functions 
${\bf{\cal K}} \equiv ({\cal K}_1, {\cal K}_2)$ is constructed from
${\cal H}_i$: 
${\cal K}_1 = \frac{{\cal H}_2 - {\cal H}_1}{{\cal H}_2 + {\cal H}_1}$,
${\cal K}_2 = \frac{{\cal H}_3 - {\cal H}_2}{{\cal H}_3 + {\cal H}_2}$,
(${\cal K}_{1,2} \le 1$).

The shape functions are given below for certain `idealized surfaces':

(1) pancake with vanishing thickness: ${\cal H}_3 \simeq
{\cal H}_2 >> {\cal H}_1$ and ${\bf{\cal K}} \simeq (1, 0)$,

(2) filament with infinitesimal diameter:
${\cal H}_3 >> {\cal H}_2 \simeq {\cal H}_1$ and
${\bf{\cal K}} \simeq (0, 1)$,

(3) sphere: ${\cal H}_3 \simeq {\cal H}_2 \simeq {\cal H}_1$ and
${\bf{\cal K}} \simeq (0, 0)$,

(4) ribbon: ${\cal H}_3 >> {\cal H}_2 >> {\cal H}_1$
and ${\bf{\cal K}} \simeq (1, 1)$.

Realistic surfaces will be represented as points on a `shape plane'
$({\cal K}_1, {\cal K}_2)$, with ideal pancakes, filaments, ribbons and spheres
defining its four vertices:
$(1, 0), (0, 1), (1, 1), (0, 0)$.

To demonstrate the effectiveness of the shape statistic we apply it to two surfaces-- an ellipsoid and a torus. The surface of the triaxial ellipsoid 
has the parametric form

\begin{equation}
{\bf{r}} = a(\sin{\theta}\cos{\phi})\hat{x} +
b(\sin{\theta}\sin{\phi})\hat{y}
+ c(\cos{\theta})\hat{z}
\end{equation}
where $0 \le \phi \le 2\pi$, $0 \le \theta \le \pi$.

In table 1 we show results for deformations of this ellipsoid. 

\def \tablerule {\noalign {\hrule}}
\begin {table}[htb]
\caption {Deformations of a triaxial ellipsoid with axis $a,b,c$. 
The shape functions
$V/S$, $S/C$, $C$ have dimensions of length, $({\cal K}_1, {\cal K}_2)$
are dimensionless.}
\vskip 0.2 true cm
\begin {tabular} {cccccc}
\tablerule
$a,b,c$ & deformation &$V/S$ & $S/C$ & $C$ & $({\cal K}_1, {\cal K}_2)$\\
\tablerule
(100, 100, 100)& sphere $\rightarrow$ filament & 100.00& 100.00 &100.00   & (0, 0)\\
(100, 80, 80) & & 85.45 & 86.12 & 86.97  & ($3.9\times10^{-3}, ~4.9\times 10^{-3}$)\\
(100, 50, 50)& & 58.51 & 61.92 &  69.01  & ($2.8\times 10^{-2}, ~5.4\times 10^{-2}$)\\
(100, 20, 20) & & 25.04 & 29.22 & 54.68 & ($7.7\times 10^{-2}, ~0.30$)\\
(100, 10, 10) & & 12.67 & 15.32 & 51.50 & ($9.5\times 10^{-2}, ~0.54$)\\
(100, 3, 3) & & 3.82 & 4.70 & 50.19 & ($0.10, ~0.83$) \\
\tablerule
(100, 100, 100)& sphere $\rightarrow$ pancake & 100.00& 100.00 &100.00 & (0,0)\\
(100, 100, 80)& & 91.99 & 92.89 & 93.63 & ($4.9\times 10^{-3}, ~3.9\times 10^{-3}$)\\
(100, 100, 50)& & 72.45 & 80.75 & 85.46 & ($5.4\times 10^{-2}, ~2.8\times 10^{-2}$)\\
(100, 100, 20) & & 36.58 & 68.45 & 79.88 & ($0.30, ~7.7\times 10^{-2}$)\\
(100, 100, 10) & & 19.42 & 65.28 & 78.90 & ($0.54, ~9.5\times 10^{-2}$)\\
(100, 100, 3) & & 5.98 & 63.88 & 78.57 & ($0.83, ~0.10$)\\
\tablerule
(100, 100, 3) & pancake $\rightarrow$ filament & 5.98 & 63.88 & 78.57 & ($0.83, ~0.10$)\\
(100, 70, 3) & & 5.97 & 52.27 & 67.32 & ($0.80, ~0.13$) \\
(100, 30, 3) & & 5.90 & 27.81 & 54.88 & ($0.65, ~0.33$) \\
(100, 10, 3) & & 5.47 & 10.79 & 50.88 & ($0.33, ~0.65$) \\
(100, 3, 3) & & 3.82 & 4.70 & 50.19 & ($0.10, ~0.83$) \\
\tablerule
\end {tabular}
\end {table}


Next consider a topologically more complicated surface, the elliptical torus
\begin{equation}
{\bf{r}} = (b + c\sin{\phi})\cos{\theta}~\hat{x} +
(b + c\sin{\phi})\sin{\theta}~\hat{y} + a(\cos{\phi})\hat{z}
\end{equation}
where $a,c < b$, $0 \le \theta, \phi < 2\pi$.
Table 2 shows shape functions for deformations of an elliptical
torus which are illustrated in Fig. 2.
(The inner tube of the torus has an elliptical cross-section with
$a$ \& $c$ being respectively, radii of curvature 
perpendicular and parallel to the plane of the 
torus; $a=c$ gives the usual circular torus.)

\def \tablerule {\noalign {\hrule}}
\begin {table}[htb]
\caption {Shape functions for an elliptical torus with axis $b,a,c$, ($b >
a,c$).}
\vskip 0.2 true cm
\label{table3}
\begin {tabular} {cccccc}
\tablerule
$b,a,c$ & Morphology & $({\cal K}_1, {\cal K}_2)$ & $V/S$ & $S/C$ & $C$ \\
\tablerule 
(100, 99, 3)& Pancake1 & ($0.90, ~2.9\times 10^{-2}$) & 7.05 & 136.89
& 144.94  \\
(100, 3, 99) & Pancake2 & ($0.88, ~0.20$) & 7.05 & 114.66 & 173.03 \\
(100, 3, 3)& Filament & ($0.14, ~0.93$) & 4.5 & 6.0 & 157.08  \\
(150, 20, 2) & Ribbon1 & ($0.70, ~0.80$) & 4.64 & 25.88 & 235.56 \\
(150, 2, 20) & Ribbon2 & ($0.70, ~0.80$) & 4.64 & 25.87 & 235.65 \\
(20, 19, 19) & Sphere-with-hole & ($0.14, ~-0.09$) & 28.5 & 38.0 & 31.42\\
\tablerule
\end {tabular}
\end {table}

\begin {figure}[ht]
\centering
\begin {minipage}[c]{6in}
\epsfxsize=5in
\epsfbox{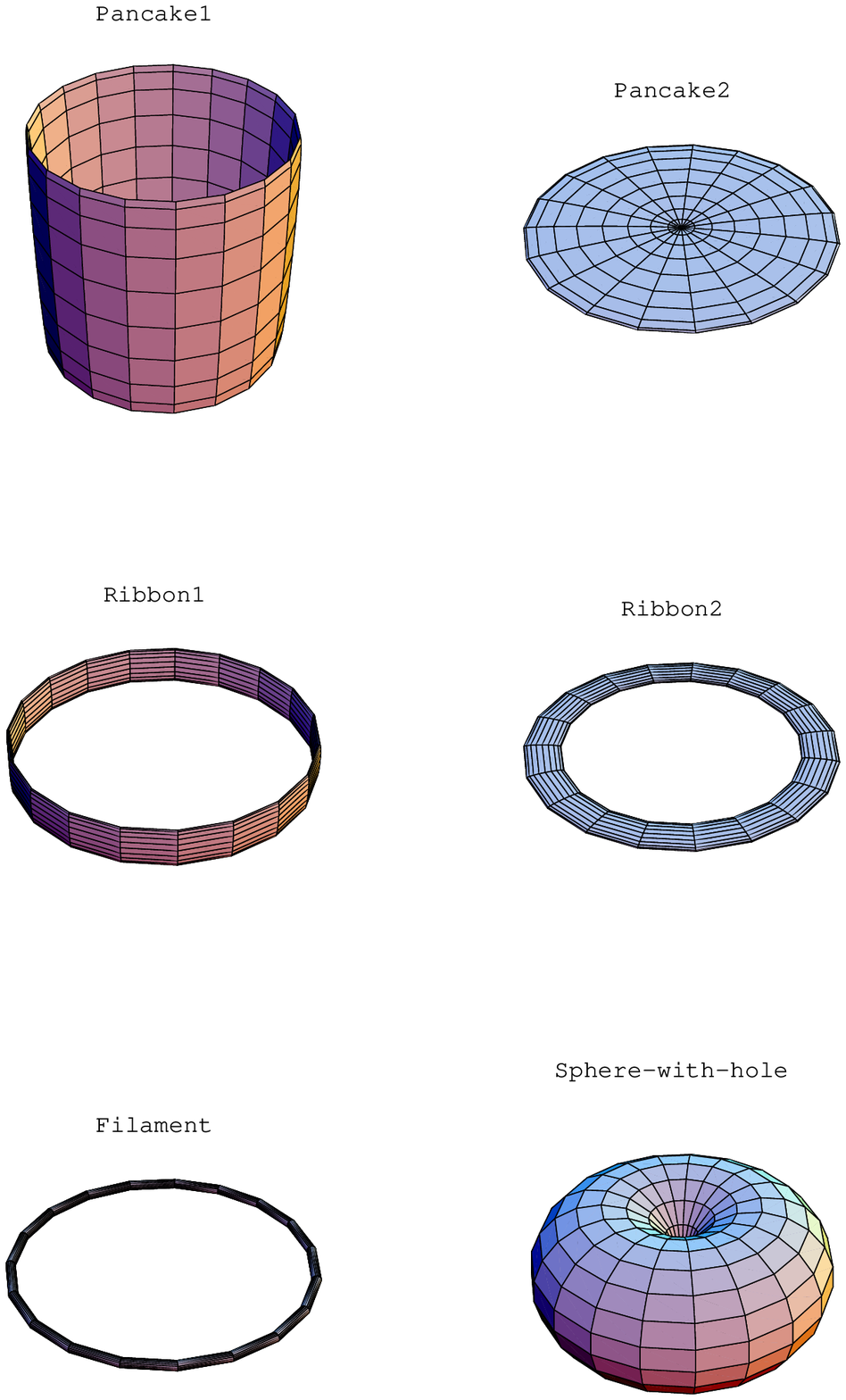}
\end {minipage}

\caption {\small Deformations of an elliptical torus.}
\end {figure}
We find that qualitatively similar deformations of the ellipsoid and torus give
rise to similar values of the shape statistic
which demonstrates that the statistic is robust.

The shape statistic we propose has several advantages over 
earlier shape-finders which were insensitive to topology \cite{sss97b,sssf97}.
As a result it can successfully be applied to 
superclusters and voids in N-body simulations and in galaxy catalogues
even when these occur at relatively low density threshold and are therefore
not necessarily simply connected. 
\footnote{The number of clusters peaks
at thresholds just below $FF_C$ making this a useful threshold at which to determine cluster shapes using our statistic.
(At $FF > FF_C$ most clusters link up to form
a percolating supercluster whereas at $FF << FF_C$ only a few clusters are 
present and these gradually disappear as $FF \rightarrow 0$.)}

\section{Conclusions.}

Non-Gaussianity in simulations of large scale
structure and in galaxy catalogs can be studied using 
geometrical
descriptors such as percolation, genus curves and shape statistics.
Geometrical indicators are sensitive to the `connectedness of a distribution and complement the two-point correlation function and the PDF.
As gravitational clustering advances, the
density field in N-body simulations
shows an increasingly pronounced departure from Gaussianity
reflected in the changing shape of the percolation curve and the
changing amplitude and shape of the genus curve, we conclude that
both genus and percolation curves provide
complementary probes of large scale structure topology
and can be used to discriminate
between models of structure formation and the analysis of observational
data such as
galaxy catalogs and MBR maps.  The smallness of the filling factor in
clusters \& superclusters at percolation indicates that a bulk of the matter
is likely to lie in filaments and pancakes. An analysis of `shapes' in N-body
simulations shows that filaments grow more pronounced as the simulation
evolves and are more prominent for spectra with greater large scale power.
To probe `shapes' more rigorously we
introduce a new shape-statistic which studies shapes of compact
surfaces (iso-density surfaces in galaxy surveys or N-body simulations)
without fitting them to ellipsoidal configurations as done earlier.
The new shape-indicators arise from simple, geometrical considerations
and are derived from fundamental properties of a compact body 
such as its
volume $V$, surface area $S$,
integrated mean curvature $C$, and Genus. 
The new shape statistics 
can be applied to topologically simple and complicated surfaces and appears to be 
quite robust.

\medskip

\noindent {\bf Acknowledgments:}
The work reported here was done in collaboration with 
B. S. Sathyaprakash and Sergei Shandarin to whom I express my gratitude and
thanks for many
years of fruitful interaction. 
I also acknowledge stimulating discussions with Sanjeev Dhurandhar and
Somak Raychaudhury.


\end {document}